\documentclass[twocolumn,showpacs,preprintnumbers,prl,amsmath,amssymb]{revtex4-1}
\usepackage{graphicx}
\usepackage{dcolumn}
\usepackage{bm}
\begin{document}
\title{Simultaneous microscopic description of nuclear level density and radiative strength function}
\author{N. Quang Hung$^{1}$}
 \email{nqhungdtu@gmail.com}
 \author{N. Dinh Dang$^{2,3}$}
  \email{dang@riken.jp}
\author{L.T. Quynh Huong$^{4,5}$}
 \affiliation{1) Institute of Research and Development, Duy Tan University, K7/25 Quang Trung, Danang City, Vietnam\\
2) Quantum Hadron Physics Laboratory, RIKEN Nishina Center for Accelerator-Based Science,
2-1 Hirosawa, Wako City, 351-0198 Saitama, Japan\\
3) Institute for Nuclear Science and Technique, Hanoi, Vietnam\\
4) Department of Natural Science and Technology, University of Khanh Hoa, Nha Trang City, Khanh Hoa Province, Vietnam\\
5) Faculty of Physics and Engineering Physics, Ho Chi Minh University of Science, Ho Chi Minh City, Vietnam}

\date{\today}
\begin{abstract}
Nuclear level density (NLD) and radiative strength function (RSF) are described simultaneously within a microscopic approach, which takes into account the thermal effects of the exact pairing as well as the giant resonances within the phonon-damping model. The good agreement between the results of calculations and experimental data extracted by the Oslo group for $^{170, 171, 172}$Yb isotopes shows the importance of exact thermal pairing in the description of NLD at low and intermediate excitation energies and invalidates the assumption based on the Brink-Axel hypothesis in the description of the RSF.
\end{abstract}

\maketitle
The rapid decrease in level spacing between the excited states as the excitation energy increases to several MeV leads to an exponential increase in the level densities and transition probabilities between the excited levels in the medium and heavy nuclei. In this condition it is impractical to deal with an individual state. Instead, it is meaningful and convenient to consider the average properties of nuclear excitations. Two main quantities, which are often employed to describe these properties, are the nuclear level density (NLD) and radiative $\gamma$-ray strength function (RSF). The NLD is defined as the number of excited levels per unit of excitation energy $E^{*}$, whereas the RSF is the average transition probability per $\gamma$-ray energy $E_\gamma$. The NLD provides the information on several properties of an atomic nucleus, namely the pairing correlations and nuclear thermodynamic properties such as temperature, entropy, heat capacity, etc.~\cite{Ericson}. The RSF reveals the characteristics of average nuclear electromagnetic properties~\cite{BW}. These two quantities have important contributions in the study of low-energy nuclear reactions and nuclear astrophysics such as the calculation of the stellar reaction rates and the description of nucleosynthesis in stars~\cite{Raus,Raus1}. The study of NLD and RSF has therefore been one of the most important topics in nuclear structure physics. It became particularly attractive after the recent developments of the experimental technique proposed by Oslo's group (the Oslo method), which is able to extract simultaneously both NLD and RSF from the primary $\gamma$-decay spectrum of the residual compound nucleus created in the transfer and/or inelastic scattering reactions~\cite{Oslo1, Oslo2, Oslo3}. These experimental data also serve as a good testing ground for all the present theoretical approaches to NLD and RSF. 

Although the concepts of NLD and RSF are rather old~\cite{Bethe,BW}, a unified theory, which can describe simultaneously and microscopically both the NLD and RSF is still absent so far. The NLD can be described quite well within the finite-temperature shell model quantum Monte-Carlo method~\cite{QMC}, but this method is time consuming when it is applied to heavy nuclei. Regarding the $\gamma$-strength functions, which involve giant resonances and the related RSF, they are beyond the scope of this method. The Hartree-Fock BCS~\cite{HFBCS} and Hartree-Fock-Bogolyubov plus combinatorial method (HFBC)~\cite{HFB} have provided a global description of NLD and might be considered as the most microscopic theories for the NLD up to date. However, they both violate the particle number. Consequently, to fit the experimental data, the NLD predicted by these theories has to be renormalized by using two parameters, whose values are extracted from the experimental analysis of the cumulative number of levels and s-wave neutron spacing at the neutron binding energy [Eq. (9) of ~\cite{HFB1}]. For those nuclei, whose experimental data are not available, the predictive power of these theories is questionable. 

Concerning the RSF, there have been few phenomenological models such as the Kadmenskij-Markushev-Furman model (KMF)~\cite{KMF} and the generalized Lorentzian (GLO)~\cite{GLO} model, and only one microscopic approach, which is the quasiparticle random-phase approximation (QRPA)~\cite{QRPA}. The KMF and GLO use several parameters such as the energy, cross section, width, centroid of $E1$, $E2$, and $M1$ resonances, whose values are found by fitting to the experimental RSF. Within the QRPA, the $\gamma$-strength function is calculated based on the normalized Lorentzian distribution, from which the resonance width and energy of the giant dipole resonance (GDR) are extracted. The $E1$-strength functions for 3,317 nuclei were extensively calculated within the QRPA and the results have been uploaded on RIPL-3 database~\cite{RIPL3}. Because the QRPA calculations were performed only for the $E1$ strengths, the results obtained within this model have not been adjusted to the experimental RSF, which consists of $E1$ as well as $E2$, and $M1$ strengths. Moreover, the predictions within the KFM and GLO models have shown that, to fit the experimental data of the RSF at the low $E_\gamma$, the width of the GDR should depend on temperature~\cite{KMF, GLO}. Since the GDR width obtained within the QRPA is temperature-independent, the predicted $\gamma$-strength functions cannot describe the experimental data unless a normalization is applied for data fitting.

It is therefore highly desirable to develop a unified microscopic theoretical approach, which can simultaneously describe both the NLD and RSF. This approach should employ only the parameters taken over from previous calculations without introducing new parameters. It has been shown that thermal pairing is crucial in the description of the NLD \cite{HungPRC81,HungPRC82} and $E1$ strength function at the excitation energies below the particle-threshold energy~\cite{DangPRC86, DangJPG40}. Moreover, as mentioned above, the temperature dependence of the GDR width is also important for the description of the RSF. In the present paper we propose, for the very first time, a theoretical approach, which takes into account both the effects of exact thermal pairing and temperature-dependent resonance width. Within our approach, thermal pairing is treated based on the eigenvalues ${\cal E}_S$, obtained by diagonalizing the pairing Hamiltonian~\cite{Volya} $H = \sum_k \epsilon_k(a_{+k}^\dagger a_{+k} + a_{-k}^\dagger a_{-k}) - G\sum_{kk'} a_{+k}^\dagger a_{-k}^\dagger a_{-k'}a_{+k'} $ at zero temperature and different numbers of unpaired particles (seniorities) $S$. Here, $a_{\pm k}^\dagger (a_{\pm})$ are the creation (annihilation) operators of a nucleon with angular momentum $k$ (in the deformed basis), projection $m_{\pm k}$, and energy $\epsilon_k$, the total seniorities $S$ are equal to $0, 2,..., \Omega$ (number of single-particle levels) for a system with an even number of particles and $1, 3,...,\Omega$ for a system with an odd number of particles. These exact eigenvalues are then used to construct the partition function of the canonical ensemble (CE) (See, e.g., Eq. (7) of Ref.~\cite{HungPRC79}). Knowing the partition function, one can easily calculate all the thermodynamic quantities such as free energy ${\cal F}$, total energy ${\cal E}$, entropy ${\cal S}$, heat capacity ${\cal C}$, and thermal pairing gap $\Delta$~\cite{HungPRC81,DangPRC86}. Because of the limitation by the size of the matrix to be diagonalized, the exact solutions of the pairing Hamiltonian are limited to the levels around the Fermi surface (truncated levels). To find the total partition function of the whole system, the exact CE partition function of the truncated levels is combined with those obtained within the independent-particle model (IPM) \cite{IPM} for the levels beyond the truncated space, where the independent motion of nucleons is assumed (that is without pairing). The total partition function is then given as the sum of the exact CE partition function for the truncated levels and the IPM partition function for the levels beyond the truncated region. The latter is obtained as the difference between the partition function of the entire single-particle spectrum (from the bottom of the single-particle potential to the closed shell $N=126$) and that of the truncated levels, for which exact pairing is taken into account~\cite{HungPRC81,HungPRC82}.

By using the inverse Laplace transformation of the partition function~\cite{Ericson}, one obtains the density of state $\omega(E^*)$ at excitation energy $E^*$ as $\omega(E^*) = {e^{\cal S}}/(T\sqrt{2\pi {\cal C}})$ \cite{Nakada99}. The total NLD $\rho(E^*)$ is obtained from the state density $\omega(E^*)$ via the relation $\rho(E^*) = {\omega(E^*)}/(\sigma\sqrt{2\pi})$ \cite{Behkami73}, where $\sigma$ is the spin cut-off parameter. 
In axially deformed nuclei, there are two spin cut-off parameters, namely
the perpendicular $\sigma_{\bot}={\cal I}_{\bot}T/\hbar^2$  and parallel $\sigma_{\|}={\cal I}_{\|}T/ \hbar^2$ ones, 
associated with the moments of inertia  perpendicular (${\cal I}_{\bot}$) 
and parallel (${\cal I}_{\|}$) to the nuclear symmetry axis. Based on the limit of rigid body with the same density distribution as of the nucleus, $\sigma_\bot$ is empirically given in the form
 $\sigma_\bot^2 \approx 0.015A^{5/3}T$ \cite{Dilg}, whereas $\sigma_{\|}$ is expressed in terms of $\sigma_\bot$ as $\sigma_{\|}=\sigma_\bot\sqrt{(3-2\beta_2)/(3+\beta_2)}$ \cite{Junghans} with $\beta_2$ and $A$ being the quadrupole deformation parameter and mass number, respectively. The collective vibrational and rotational excitations, not included in the pairing Hamiltonian, also significantly increase the NLD. These increases are expressed in terms of the vibrational $k_{\text{vib}}$ and  rotational $k_{\text{rot}}$ enhancement factors, defined as the ratio between the ``correct" NLD including all degrees of freedom and the NLD where the collective vibration and rotation are respectively absent~\cite{Junghans,KvKr,Kvib}. Their explicit forms are given based on the empirical formulas as 
 $k_{\text{vib}}={\exp}[0.0555 A^{2/3}T^{4/3}]$~\cite{Kvib} and $k_{\text{rot}}=(\sigma_\bot^2-1)/[1+e^{(E^*-U_C)/D_C}]+1$, where $E^*$ is the excitation energy obtained within the exact CE of the pairing Hamiltonian plus the IPM (EP+IPM), whereas $D_C$ and $U_C$ are given as $D_C=1400\beta_2^2 A^{-2/3}, U_C=120\beta_2^2 A^{1/3}$~\cite{Junghans}.  An alternative treatment of $k_{\text{vib}}$ based on the generalized boson partition function has been reported in Ref. \cite{HFB1}, where the coherent particle-hole ($ph$) configurations forming the collective phonons are separated from the incoherent ones to avoid double counting. The distribution of $k_{{\rm vib}}$ found in this way in the region of $E^* < $ 30 MeV is quantitatively equivalent to the empirical formula used in the present paper. The final total NLD, including the effects of vibrational and rotational enhancements, is given as $\rho(E^*) = k_{\text{rot}}k_{\text{vib}}{\omega(E^*)}/{(\sigma_{\|}\sqrt{2\pi})}$ \cite{Junghans,RIPL}.

The RSF $f_{X\lambda}(E_\gamma)$ for the electric ($X=E$) or magnetic ($X=M$) excitations with multipolarity $\lambda$ is calculated via the $X\lambda$ strength function $S_{X\lambda}(E_\gamma)$. In the phenomenological models, a Lorentzian is used for the strength function 
$S_{X\lambda}(E_\gamma)$ with an approximated resonance width for $E1$ excitations (the KMF) as a function of $T^2$, whereas the widths for $M1$ and $E2$ excitations take their values at $T=$ 0 as $T$ varies (See Eqs. (14) -- (17) in Ref. \cite{Oslo2} or Eqs. (9) -- (11) in Ref. \cite{Voinov} ). These assumptions are generally incorrect because the giant resonance widths are known to be temperature-dependent, but the $T^2$ dependence of the $E1$ resonance width, which the KMF borrows from the collisional damping model, is a good approximation only up to $T\sim$ 1 MeV (See Fig. 10 in Ref. \cite{EPJA}). Moreover, the effect of thermal pairing at low $T$ was completely neglected in these phenomenological models. 

In the present work, we calculate the strength function $S_{X\lambda}(E_\gamma)$ within the Phonon Damping Model (PDM), where the temperature-dependent resonance width $\Gamma_{X\lambda}(T)$ is obtained microscopically, including the effect of non-vanishing thermal pairing~\cite{PDM}. Moreover, instead of using the approximate pairing as in Ref. \cite{PDM}, we employ the exact CE pairing mentioned above. The formalism of the PDM with exact pairing at zero and finite temperatures has been reported in Ref. \cite{DangJPG40}. The PDM has also been discussed in a series of papers, whose most recent review is given in Sec. 3.5 of \cite{EPJA}. The resonance width in the PDM is the sum of the quantal width $\Gamma_Q$ caused by coupling the collective giant excitations to the non-collective $ph$ configurations at zero and non zero $T$, and the thermal width $\Gamma_T$ caused by coupling of giant resonances to $pp$ and $hh$ configurations at $T\neq$ 0 (See Eqs. (1a) -- (1c) in \cite{PDM}). The model has two parameters $F_1^{(\lambda)}$ and $F_2^{(\lambda)}$ for the couplings to $ph$, and $pp$ ($hh$) configurations, respectively. The value of $F_1^{(\lambda)}$ is chosen to reproduce the resonance width $\Gamma_{X\lambda}(T=0)$, whereas $F_2^{(\lambda)}$ is selected at $T=$ 0 so that the resonance energy $E_{X\lambda}$ does not changes significantly as $T$ varies. In numerical calculations in the present work, the small fluctuation of the resonance peak is neglected by setting the resonance energies $E_{X\lambda}$ for $E1, M1$, and $E2$ excitations at their corresponding experimental value extracted at $T=$ 0. The numerical calculations are carried out for $^{170, 171, 172}$Yb isotopes, whose single-particle spectra are taken from the axially deformed Woods-Saxon potential \cite{Cwoik}. The quadrupole deformation parameters $\beta_2$ are 0.295 for $^{170, 171}$Yb and 0.296 for $^{172}$Yb, whereas other parameters of the Woods-Saxon potential are the same as those reported in Refs. \cite{HungPRC81,HungPRC82}. The values of the pairing interaction parameter $G$ for neutrons and protons are chosen so that the exact neutron and proton pairing gaps obtained at $T=0$ reproduce the corresponding experimental values extracted from the odd-even mass formulas. The diagonalization of the pairing Hamiltonian is carried out for 12 doubly degenerate single-particle levels with 6 levels above and 6 levels below the Fermi surface. A set of total 73,789 eigenstates for each type of particles is obtained and employed to construct the exact CE partition function. By using Eqs. (11) and (12) of Ref. \cite{DangPRC86}, the exact CE chemical potential and pairing gap are calculated, from which one obtains the quantities that mimic the ``exact" quasiparticle energy $E_k$, the coefficients $u_k$ and $v_k$ of the Bogolyubov transformation between particles and quasiparticles, as well as the quasiparticle occupation numbers $n_k$ based on their conventional definitions (See, e.g., Eqs. (3), (4), and (13) of Ref. \cite{DangPRC86}). These quantities are used as inputs in the RSF calculations within the PDM for the levels with pairing around the Fermi surface, whereas for the remaining spectrum, where $u_k=$ 1 (0) and $v_k=$ 0 (1) for $k= p$ $(h)$ according to the IPM, one has $E_k = |\epsilon_k - \epsilon_F|$, $n_p = f_p$ and $n_h = 1-f_h$ with $\epsilon_F$ and $f_k$ being the Fermi energy and the single-particle occupation number described by the Fermi-Dirac distribution at finite $T$, respectively.

\begin{figure}[h]
\begin{center}
\includegraphics[width=8.7cm]{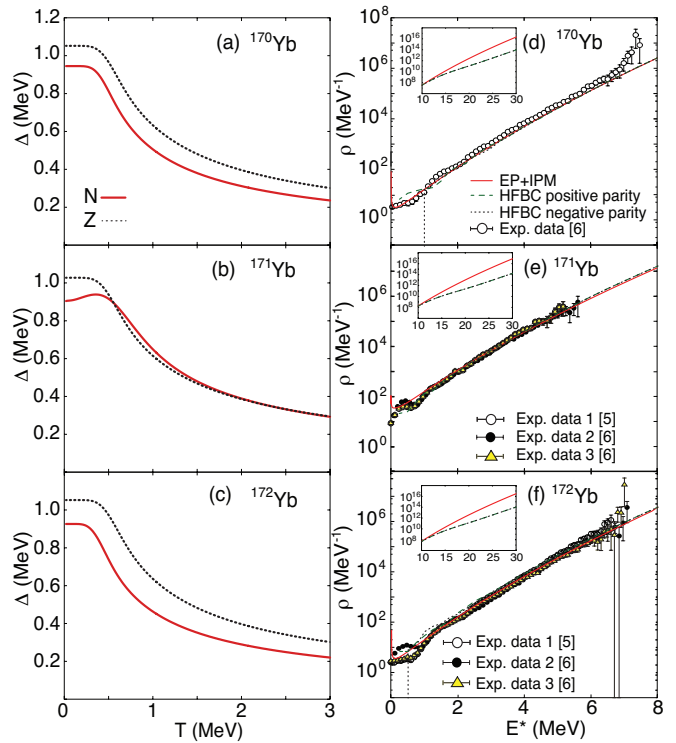}
\caption{(Color online) Neutron and proton pairing gaps $\Delta$ [(a) -- (c)] as functions of $T$ and total level densities $\rho$ [(d) -- (f)] as functions of $E^{*}$ obtained within the EP+IPM in comparison with predictions of HFBC calculations for the positive and negative parities and the experimental data for $^{170, 171, 172}$Yb nuclei.}
\label{fig1}
\end{center}
\end{figure}
The results of the exact neutron (solid lines) and proton (dotted lines) gaps as functions of $T$ are plotted in Figs. \ref{fig1}(a) -- \ref{fig1}(c). It is seen in these figures that the exact gaps decrease with increasing $T$ and remain finite up to $T$ as large as 3 MeV, well above the critical temperature $T_c\sim0.57\Delta(T=0)$, where the pairing gap collapses within the approximate theories such as the BCS one. A slight increase in the exact neutron gap at low $T < 0.5$ MeV  is seen in $^{171}$Yb because of the blocking effect from the odd neutron~\cite{HungPRC94}. Owing to this non-vanishing pairing gaps, the NLDs obtained within the EP+IPM (solid lines) agree well with the experimental data for all nuclei considered in the present paper as seen in Figs. \ref{fig1}(d) -- \ref{fig1}(f). These panels also show that the NLDs obtained within the EP+IPM almost coincide with results of the global microscopic calculations within the HFBC for both negative (dashed lines) and positive (dotted lines) parities, whose values are taken from the RIPL-3 database~\cite{RIPL3}. However, as has been mentioned above, to have a good description of the experimental data the NLDs obtained within the HFBC have to be renormalized based on two phenomenological parameters, spoiling their microscopic nature. Moreover, since the HFBC was derived based on the partition function of the incoherent $ph$ states built on top of the HFB single-particle spectra, it is certainly not able to predict the NLD in the region of high excitation energy, where the contributions of the $pp$, $hh$, as well as of higher states like $2p2h, 3p3h$, etc. become significant. Meanwhile, within the EP+IPM, the exact CE partition function is obtained from the direct diagonalization of the matrix elements of the Hamiltonian, which consist of all possible couplings between the $ph$, $pp$ and $hh$ states. Therefore, this exact CE partition, when combined with that of the IPM, is capable to describe the NLD up to high $E^*$ region. The insets of Figs. \ref{fig1}(d) -- \ref{fig1}(f), where the NLDs obtained within the EP+IPM are compared with those obtained within the HFBC in the region 10 $\leq E^*\leq$ 30 MeV, clearly show that the former are significantly higher than the latter. The merit of the EP+IPM is also in the fact that, beyond Woods-Saxon mean field, it uses only two parameters, namely the monopole pairing strength parameters $G$ for protons and neutrons, which are adjusted to fit the corresponding experimental gaps at $T=0$.

\begin{figure}[h]
\begin{center}
\includegraphics[width=8.7cm]{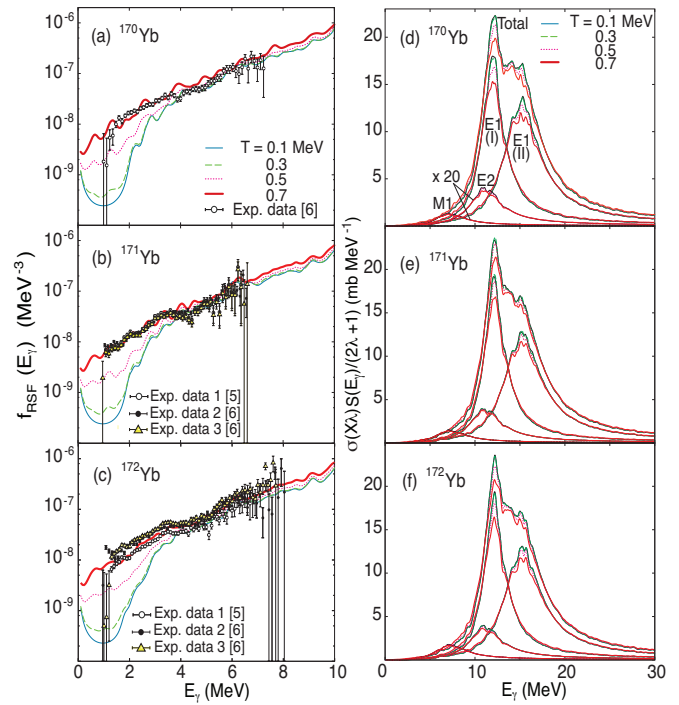}
\caption{(Color online) Radiative strength functions [(a) -- (c)] obtained within the EP+PDM in comparison with experimental data for $^{170, 171, 172}$Yb nuclei, and the corresponding total strength functions [(d) -- (f)] together with their components for $E1$, $E2$, and $M1$ excitations as functions of $E_{\gamma}$ at different temperatures.}
\label{fig2}
\end{center}
\end{figure}
Shown in Fig. \ref{fig2} are the RSF [(a) -- (c)] and the sum ${\cal S}_{PDM}(E_{\gamma})$ of the strength functions $S_{X\lambda}(E_{\gamma})$ calculated within the PDM for $E1$, $M1$, and $E2$ resonances at several values of $T\leq$ 0.7 MeV [(d) -- (f)]. These strength functions have been multiplied by the corresponding cross sections $\sigma(X\lambda)$ at their maxima and normalized by $(2\lambda+1)$, namely ${\cal S}_{PDM}(E_{\gamma})=\sigma(E1(I))S_{E1(I)}(E_{\gamma})/3+\sigma(E1(II))S_{E1(II)}(E_{\gamma})/3+\sigma(M1)S_{M1}(E_{\gamma})/3+\sigma(E2)S_{E2}(E_{\gamma})/5$, where $E1(I)$ and $E1(II)$ correspond to the two components of the GDR determined from the photoabsorption experiments~\cite{Oslo2}. The values of resonance energies $E_{X\lambda}$, their FWHM $\Gamma_{X\lambda}$, and cross sections $\sigma(X\lambda)$ at $T=$ 0 for $^{170, 171, 172}$Yb are taken from Table I of Ref. \cite{Oslo2}. The GDR with the largest values of $\sigma(X\lambda)$ ($X\lambda = E1(I), E1(II)$) gives the largest contribution the total strength function [Figs. \ref{fig2}(d) -- \ref{fig2}(f)]. The widths of its two components remain nearly constant at $T \leq$ 0.4 MeV and increase with $T$ at $T>$ 0.4 MeV, resulting in a significant increase in the total RSF at low $E_\gamma < 4$ MeV as seen in Figs. \ref{fig2}(a) -- \ref{fig2}(c). The RSFs obtained within the PDM at $T=$ 0.7 MeV agree well with the experimental data for all nuclei under consideration. This value of $T$ is higher than that obtained from the fitting by using the KFM model in Ref. \cite{Oslo2}, which is always below 0.4 MeV. This result is very important as it invalidates the assumption of the Brink-Axel hypothesis~\cite{BA}, which states that the GDR built on an excited state should be the same as that built on the ground state, and based on which the experimental RSFs were extracted. Based on the fitting by using the KMF model, Ref. \cite{Oslo2} has also suggested that there should appear a two-component pygmy dipole resonance (PDR) in the region 2.1 $<E_{\gamma}<$ 3.5 MeV in $^{171}$Yb and $^{172}$Yb. Although not reproduced in any microscopic models so far, this two-component PDR was added on top of the GDR in fitting the experimental RSF in Ref. \cite{Oslo2}. Within the PDM, it has been found in Ref. \cite{DangJPG40} that exact pairing enhances the $E1$ strength function in the region $E_\gamma<$ 5 MeV. Including this exact pairing, which does not vanish at $T>T_c$, the RSFs, calculated within the PDM, agree well with the experimental data [thick solid lines in Figs. \ref{fig2}(e) and \ref{fig2}(f)]. In this way, the enhancement of the experimental RSF at low $E_\gamma$, which was suggested to be caused by the PDR, is explained microscopically by the effect of exact thermal pairing within the PDM.

In conclusion, we propose for the very first time a microscopic approach, which is able to describe simultaneously the nuclear level density and radiative $\gamma$-ray strength function. This approach used the exact solutions of the pairing problem to construct the partition function to calculate the NLD and thermal pairing gap at finite temperature. The latter is included in the PDM to calculate the RSF. The good agreement between the results obtained within this approach and the experimental data for NLD and RSF in $^{170, 171, 172}$Yb has shown that exact thermal pairing is indeed very important for the description of both NLD and RSF in the low and intermediate region of excitation and $\gamma$-ray energies. Moreover, to have a good description of the RSF the microscopic strength function with the temperature-dependent width for the giant resonances should be used instead of the Brink-Axel hypothesis. The merits of this approach are its microscopic nature and the use of only the parameters taken over from previous calculations. It does not consume much computing time either as the calculation takes less than five minutes even for a heavy nucleus, and therefore can be performed on a PC.
\acknowledgments
The numerical calculations were carried out using the FORTRAN IMSL Library by Visual Numerics on the RIKEN supercomputer HOKUSAI-GreatWave  System.

\end{document}